%% file: main.tex
\documentclass[a4paper,11pt]{article}
\pdfoutput=1 %
\usepackage{jcappub}
\usepackage{booktabs}  %
\usepackage{multirow}
\usepackage{arydshln}
\usepackage[normalem]{ulem}
\usepackage[T1]{fontenc}
\usepackage{orcidlink}

\title{Cosmological constraints from the DESI DR1 joint power spectrum and bispectrum analysis}

\input{authorlist}
\emailAdd{sergi.novell@icc.ub.edu, hectorgil@icc.ub.edu, liciaverde@icc.ub.edu}

\abstract{We derive cosmological parameter constraints from the Dark Energy Spectroscopic Instrument (DESI) Data Release 1 (DR1) galaxy clustering data, based on a joint full-shape analysis of the power spectrum multipoles and the bispectrum monopole using the ShapeFit framework \cite{brieden2021shapefit}. This work is the follow-up of \cite{novellmasot25fullshape}, which obtains for the first time constraints on the ShapeFit parameters using the bispectrum of DESI DR1, for the luminous red galaxy (LRG) and quasar (QSO) samples. Here we present the first ShapeFit cosmological inference results using the bispectrum of DESI DR1. For the baseline flat $\Lambda$CDM model, our analysis yields results broadly in agreement with previous studies of the DESI DR1 data-set. In particular, we recover values for the matter density parameter and Hubble constant of respectively $\Omega_m=0.310\pm0.012$ and $H_0=[68.92\pm0.97]\,\mathrm{km\, s^{-1} Mpc^{-1}}$, consistent with previous results from the full DESI DR1 dataset that did not use the bispectrum signal~\cite{KP5Desi}. The inclusion of the bispectrum significantly tightens the constraints on the amplitude of fluctuations, reducing the error-bars in $\ln(A_s\times10^{10})$ by approximately 20\%, compared to using the power spectrum alone. We also explore extended cosmological models by performing fits for the evolving dark energy equation of state $w_0w_a$, and the sum of neutrino masses $\sum m_\nu$. 
In these cases, we obtain constraints slightly larger than the ones from previous works from the DESI collaboration \cite{KP7}, due to not combining the full-shape results with other probes in all tracers. We find no strong evidence of deviations from standard $\Lambda$CDM, with the dark energy equation-of-state remaining within 2$\sigma$ from a cosmological constant $\Lambda$, and the neutrino mass being consistent with the normal hierarchy, $\sum m_\nu<0.1\,[eV]$ at 95\% confidence limit.
These constraints are broadly consistent with other DESI DR1 analyses, thus validating the robustness of the ShapeFit compression approach and the inclusion of the bispectrum for cosmological inference.}

\begin{document}
\maketitle
\flushbottom
\section{Introduction}\label{sec:intro}
Increasingly large and high-precision cosmological datasets are allowing for even more stringent tests of the cosmological model.
The $\Lambda$-Cold Dark Matter ($\Lambda$CDM) framework has been remarkably successful at explaining observations across cosmic history \cite{collaboration2018planck}. Yet, challenges remain, %
including the unknown nature of the dark components (dark matter and dark energy) that dominate the energy density of the Universe, as well as observed tensions between early- and late-Universe measurements \cite{verde2019tensions,kp6bao,KP7}. These persistent issues motivate the continued exploration of potential new physics beyond $\Lambda$CDM.

Galaxy redshift surveys have become one of the primary ways of probing the large-scale structure (LSS) of the Universe. Surveys such as the Sloan Digital Sky Survey (SDSS) \cite{gunn20062}, the 2-degree Field Galaxy Redshift Survey (2dfGRS) \cite{percival20012df}, the Baryon Oscillation Spectroscopic Survey (BOSS) \cite{dawson2012baryon}, and the extended BOSS (eBOSS) \cite{dawson2016sdss} successfully obtained LSS cosmological constraints from increasingly larger datasets.

Building on this, the Dark Energy Spectroscopic Instrument (DESI) \cite{Snowmass2013.Levi,DESI2016a.Science,DESI2016b.Instr,DESI2022.KP1.Instr,FocalPlane.Silber.2023,Corrector.Miller.2023,Spectro.Pipeline.Guy.2023,SurveyOps.Schlafly.2023,LRG.TS.Zhou.2023,DESI2023a.KP1.SV,DESI2023b.KP1.EDR,kp3bao,kp4lya,adame2024desi,kp6bao,FiberSystem.Poppett.2024,kp2,ross2025construction}
is  obtaining spectra for more than 60 million objects \cite{Levi:2013gra}, including luminous red galaxies (LRG) \cite{LRG.TS.Zhou.2023}, emission-line galaxies (ELG) \cite{elgdesi}, and quasars (QSO) \cite{qsodesi}, as well as the nearby bright galaxy survey (BGS) \cite{bgsdesi} and the high redshift Lyman-$\alpha$ forest \cite{kp4lya}. DESI is mapping a cosmological volume many times larger than all previous surveys combined, which enables exquisite precision in the measurement of the traditional signatures in LSS cosmology, namely, baryon acoustic oscillation (BAO) and redshift-space distortions (RSD). The density of tracers measured by DESI provides enough statistical power to perform analyses beyond two-point, which is key to break parameter degeneracies, %
tighten cosmological constraints and provide additional tests for $\Lambda$CDM \cite{Verde1998,Matarrese1997,Sefusatti:2006pa,fry1993biasing,Hahn_2020,oddo2021cosmological,bartolo2013matter,bellini2015signatures,bertacca2018relativistic,Gil_Mar_n_2011,yankelevich2019cosmological,coulton2019constraining,ruggeri2018demnuni,gagrani2017information,Gualdi:2020aniso,rizzo2023halo,ivanov2023cosmology,d2022one, baldauf2011primordial,dizgah2021primordial, gualdi_matter_2021}. Indeed, even though higher-order statistics such as the bispectrum or three-point correlation function have been long studied,  only a few works applied them to galaxy surveys \cite{Verde:2001pf, Scoccimarro01, gil-marin_power_2015, HGMbispectrum15b, gil-marin_clustering_2017,ivanov2023cosmology, philcox2022boss,d2022boss,d2022limits,gualdi_geometrical_2019,Gualdi:2018pyw,cabass2022constraints,cabass2022constraintsmulti}.

The information contained in the BAO and RSD signals of the galaxy catalogue summary statistics (such as the power spectrum or the bispectrum) constitutes only {\it a partial} set of the whole information contained in those statistics. 
Standard BAO and RSD analyses marginalize over features present in the broadband shape of $P(k)$, focusing on two robust observables: the BAO scale --often parametrized by the dilation factors $\alpha_\parallel,\alpha_\bot$ along and across the line of sight, respectively-- and the product of the logarithmic growth rate $f$ and the averaged amplitude of perturbations $\sigma_8$. While this \textit{compressed} approach has the virtue of being largely model-independent, it is not able to capture the full information enclosed in the power spectrum. One way to do so is through the so-called Full Modelling analysis of the power spectrum (sometimes called ``direct fit''), which fits the power spectrum directly from the theoretical predictions for any given cosmology. %
 Such an approach yields tighter constraints by exploiting information enclosed in the broadband shape of the matter power spectrum. However, the most popular implementation of such analysis requires adopting informative priors over poorly known nuisance parameters.

The ShapeFit methodology \cite{brieden2021shapefit,brieden2021letter, Brieden_ptchallenge, brieden2022model} was proposed as a middle ground between the standard BAO-RSD compression and a Full Modelling analysis. In a ShapeFit-based analysis, one still performs a template fit to the galaxy power spectrum, but includes two additional parameters ($m,n$) that model the slope of the power spectrum near the matter-radiation equality scale. After obtaining constraints for the full set of ShapeFit parameters and performing cosmological inference on them, one can recover results with statistical power close to those of direct full-shape fits for models close to $\Lambda$CDM \cite{brieden2021letter} and its popular extensions \cite{maus2024comparison,maus2024analysis, noriega2024comparing, lai2024comparison, ramirez2024full}.

In our previous work \cite{novellmasot25fullshape}, we applied the ShapeFit compression technique to the clustering data from DESI DR1. In particular, we constrained the ShapeFit parameters using the power spectrum and bispectrum statistics in the LRG and QSO tracers, spanning redshifts between $0.4<z<2.5$. We obtained joint power spectrum-bispectrum results highly compatible with the power spectrum-only analysis from the DESI collaboration \cite{KP5Desi}, with the addition of the bispectrum yielding tighter constraints, as well as disentangling the $f$ and $\sigma_8$ parameters.

In this work, we use the results from that analysis to infer cosmological constraints under various model assumptions. Because the ShapeFit compression is model-agnostic, constraints on the compressed variables can simply be re-interpreted within the chosen cosmological model. In particular, we start by obtaining the $\Lambda$CDM results from the ShapeFit constraints with and without the inclusion of the bispectrum. We also compare our main approach (using power spectrum and bispectrum) with the analogous case using the traditional BAO+RSD results,  to highlight the effect of adding the shape parameter in the analysis. 

Then, we obtain sound horizon-free constraints for the $\Omega_\textrm{m}$ and $H_0$ parameters, in which we interpret the BAO signal without assuming the sound-horizon scale $r_\textrm{d}$ from early-time physics. We also explore other relevant cosmological models,  constraining the time-varying dark energy equation of state (using the Chevallier-Polarski-Linder (CPL) parametrization with $w_0$ and $w_a$ \cite{chevallier2001accelerating,lindercpl}), and the total neutrino mass $\sum m_\nu$.%

This paper is organized as follows. In Section \ref{sec:methodology} we describe the data and the ShapeFit methodology and measurements, including details of the ShapeFit implementation and the cosmological parameter inference procedure. Section \ref{sec:results} presents our results: we first discuss the baseline $\Lambda$CDM constraints (Section \ref{sec:baseline}). We then report the findings for each extended scenario: sound-horizon marginalization (Section \ref{sec:shfree}), evolving dark energy (Section \ref{sec:w0wa}), and neutrino mass (Section \ref{sec:mnu}).%
In Section \ref{sec:discussion}, we discuss the implications of our results and conclude.

\section{Methodology}\label{sec:methodology}

\subsection{ShapeFit compression.}
Following the ShapeFit methodology described in \cite{brieden2021shapefit,brieden2022model}, rather than directly performing a direct fit of a cosmological model to the measured power spectra and bispectra, we first compress the data into a set of intermediate compressed parameters. The full details of this procedure and parameters can be found in \cite{novellmasot25fullshape}. In short, for each galaxy sample and redshift bin, we fit the joint power spectrum-bispectrum data-vector to a template based on a reference (or fiducial) cosmology, obtaining constraints for the following parameters (see appendix \ref{sec:appB} for a concise summary on the ShapeFit variables and their connection to the cosmological variables):
\begin{itemize}
    \item $\alpha_\textrm{iso}(z)$: the isotropic BAO scaling, defined as the ratio of $D_V(z)/r_\textrm{d}$ relative to the fiducial or reference cosmology, $\alpha_\textrm{iso}(z)=[D_V(z)/r_\textrm{d}]/[D_V(z)/r_\textrm{d}]^{\rm ref} $ . The volume distance $D_V(z)$ is appropriately defined as $D_V(z)=(zD_M(z)^2D_H(z))^{1/3}$, with $D_M(z)$ the angular diameter distance and $D_H(z)\equiv c/H(z)$ the Hubble distance. Hence, $\alpha_\textrm{iso}(z)$ encodes information from both late-time (via $D_V(z)$) and early-time (via $r_\textrm{d}$) physics.  
    \item $\alpha_\textrm{AP}(z)$: the anisotropic Alcock-Paczynski distortion parameter, sensitive to the ratio of the Hubble distance and the angular diameter distance: \sloppy$\alpha_\textrm{AP}(z)=[D_H(z)/D_M(z)]/[D_H(z)/D_M(z)]^{\rm ref}$, which is therefore a purely late-time quantity.

    \item $\sigma_{\textrm s8}(z)$: The amplitude of matter fluctuations at redshift $z$, smoothed on the scale of $8 r_d^\textrm{ref}r_d^{-1}\textrm{Mpc}h^{-1}$. Note the change from the usual name $\sigma_8$ to $\sigma_\textrm{s8}$: In the case of $\sigma_8$, the smoothing is on the constant scale $8\,\textrm{Mpc}h^{-1}$ (making it dependent on dilations of scale, which are parametrized by $\alpha_\textrm{iso}(z),\,\alpha_\textrm{AP}(z)$)--more details in Section 3 of \cite{brieden2021shapefit}.
    
    \item $f(z)$: The logarithmic growth rate of structure. This quantity, in a $\Lambda$CDM cosmology under the assumption of general relativity, is uniquely determined by $\Omega_m$, as $f(z)=\Omega_m(z)^{0.55}$. In non-$\Lambda$CDM, but still GR,  models $f(z)$ is uniquely determined by $\Omega_m(z)$ and background quantities, while on non-GR models, this is not the case. 
    
    In two-point analyses, $f(z)$ is highly degenerate with $\sigma_{s8}(z)$, as the amplitude of two-point clustering is proportional to the product $f\sigma_{s8}(z)$. The addition of the bispectrum statistic allows constraining $f(z)$ separately by breaking the degeneracy.
    \item $m(z)+n(z)$: the combined ShapeFit parameter, obtained as the sum of two different parameters affecting the slope of the power spectrum. The parameter $m(z)$ governs the slope of the power spectrum around the matter-radiation equality scale, while $n(z)$ models variations of the primordial spectral index. These two parameters are highly degenerate for current data sets, and using their sum (which is equivalent to $m$-at-fixed-$n$, or vice versa) compresses the information of both parameters without perceptible loss, as reported in figure 9 of \cite{gsponer2025fiducial}. 
\end{itemize}

When only $\{\alpha_{\rm iso}, \alpha_{\rm AP}, f, \sigma_{s8}\}$ are employed, we refer to the analysis as a BAO+RSD (or Standard Compression), whereas when $m+n$ is also added, we refer to it as a ShapeFit analysis. In appendix~\ref{sec:compression} we display how the cosmological inference for $\Lambda$CDM for the dataset described in Table~\ref{tab:tracers} compares when using the full set of variables above (ShapeFit), or when using a subset of it: {\it i)} shape variables $m+n$ only; and {\it ii)} $\alpha_{\rm iso}$, $\alpha_{\rm AP}$ and $f\sigma_{\rm s8}$ mimicking the traditional compression analysis). 

The pipeline used to constrain the compressed parameters is calibrated exclusively on blinded data \cite{novell2024blinding,brieden2020blind,andrade2024validating}. It accounts for the effects of the survey window \cite{HGMeboss} and incorporates the full systematic error budget into the reported uncertainties.

The power spectrum model is based on two-loop renormalized perturbation theory (RPT) \cite{Crocce_2006,gil-marin_power_2015}, the bispectrum is modelled with GEO-FPT \cite{novell2023geofpt}, and the covariance matrices are estimated from large sets of mock catalogs \cite{maksimova2021abacussummit,novell2023approximations}. Therefore, we fully account for "real-world" effects, including the cross-correlation terms between the power spectrum and the bispectrum.  More details can be found in \cite{novellmasot25fullshape,novell2023approximations}.

\subsection{Data and measurements.}
The DESI DR1 is divided into several redshift bins, according to the corresponding tracer: the bright galaxy survey, BGS \cite{bgsdesi}; the luminous red galaxies \cite{LRG.TS.Zhou.2023}, LRG, which are split in the three redshift bins LRG1, LRG2, LRG3; emission line galaxies \cite{elgdesi}, ELG, divided in two redshift bins ELG1 and ELG2; and quasars (QSO) \cite{qsodesi}, as well as the Lyman-$\alpha$ forest data \cite{kp4lya}. Summary information about these tracers can be found in Table \ref{tab:tracers}.

\begin{table}[ht!]
\centering
\centering
        \small
    \resizebox{\columnwidth}{!}{
\begin{tabular}{|c|c|c|c|c|c|c|}
\hline
Tracer & Redshift range & $N_\text{tracer}$ & $z_\text{eff}$ & $P_0 \, [(\text{Mpc}h^{-1})^3]$ & $V_\text{eff} \, [\text{Gpc}^3]$ & Used in this work as \\ \hline\hline
BGS  & $0.1 - 0.4$ & 300,017   & 0.295 & $\sim 9.2 \times 10^3$ & 1.7 &  BAO recon. \\ \hline
LRG1 & $0.4 - 0.6$ & 506,905   & 0.510 & $\sim 8.9 \times 10^3$ & 2.6 &  FS \\ \hline
LRG2 & $0.6 - 0.8$ & 771,875   & 0.706 & $\sim 8.9 \times 10^3$ & 4.0 & FS \\ \hline
LRG3 & $0.8 - 1.1$ & 859,824   & 0.920 & $\sim 8.4 \times 10^3$ & 5.0 &  FS \\ \hline
ELG1 & $0.8 - 1.1$ & 1,016,340 & 0.955 & $\sim 2.6 \times 10^3$ & 2.0 &  BAO recon. \\ \hline
ELG2 & $1.1 - 1.6$ & 1,415,687 & 1.317 & $\sim 2.9 \times 10^3$ & 2.7 & BAO recon. \\ \hline
QSO  & $0.8 - 2.1$ & 856,652   & 1.491 & $\sim 5.0 \times 10^3$ & 1.5 &  FS \\ \hline
\end{tabular}
}
\caption{Main properties of the DESI DR1 tracers: redshift range, counts per redshift bin, effective redshift $z_\textrm{eff}$, power spectrum monopole amplitude at the reference scale of $k=0.14\,h\textrm{Mpc}^{-1}$, $P_0$, and effective volume $V_\textrm{eff}$. We use the full-shape results from the joint power spectrum and bispectrum analysis of \cite{novellmasot25fullshape} for the three LRG bins and the QSOs, while using reconstructed BAO constraints for the BGS and the ELG samples from \cite{kp3bao}. Additionally, we also use the Lyman-$\alpha$ BAO measurements from \cite{kp4lya}.}
\label{tab:tracers}
\end{table}

In this work, we utilise the ShapeFit constraints for the three LRG bins and the QSOs obtained by using the joint power spectrum and bispectrum analysis performed in \cite{novellmasot25fullshape}. These constraints were obtained by using the power spectrum multipoles and bispectrum monopole, where in the LRG bins the power spectrum multipoles include monopole, quadrupole and hexadecapole. For the QSO tracers, we did not use the power spectrum hexadecapole due to the lower signal present in that case (see figure 13 of \cite{novellmasot25fullshape} to show how the inclusion of the hexadecapole does not significantly improve the constraining power for the QSO sample). Ref~\cite{novellmasot25fullshape} does not perform a bispectrum analysis on the ELG sample because of large imaging systematics and the difficulty in correcting them at the level of the bispectrum signal; the BGS bispectrum analysis was not performed due to its lower effective volume and limited constraining power.  Additionally, the DR1 BAO reconstruction signal is only employed in those bins where we do not have a ShapeFit power spectrum and bispectrum measurement (ELG and BGS). Although we could have employed the DR1 BAO reconstruction signal as well for the LRG and QSO bins, where we have a power spectrum and bispectrum ShapeFit measurements, we leave that for future work in order to avoid the complication of determining the exact covariance between BAO and ShapeFit variables \cite{Forero-sanchez-novell-masotinprep}.

All ShapeFit constraints used in this work include the corresponding systematic error budget.

The data are binned in wavenumber up to a maximum $k_\textrm{max}$, defined to correspond to mildly non-linear scales, as to retain non-linear information while excluding scales where modelling systematics become significant.
We set for our analysis $k_\textrm{max}=0.15\,h\,\textrm{Mpc}^{-1}$ for the power spectrum\footnote{This is a conservative choice, which, however, is not expected to lose significant information. As seen in Fig. 20 of Ref. ~\cite{KP5Desi}, the recovered constraints are largely insensitive to the choice of $k_\textrm{max}$ within the range from $0.16-0.20\,h\,\textrm{Mpc}^{-1}$.} and $k_\textrm{max}=0.12\,h\,\textrm{Mpc}^{-1}$ for the bispectrum, the latter being the range at which the GEO-FPT model was calibrated \cite{novell2023geofpt}. 

We complement these ShapeFit measurements by adding the reconstructed BAO constraints for the remaining redshift bins, e.g. using the $\alpha_\textrm{iso}(z),\,\alpha_\textrm{AP}(z)$ measurements and covariances from \cite{kp3bao} for the BGS and ELG tracers, and from \cite{kp4lya} for the Lyman-$\alpha$ forest.

\subsection{Cosmological parameter inference.}
The compressed parameters and their covariance, both obtained in \cite{novellmasot25fullshape}, are then used to infer cosmological model parameters. The procedure is the following:
\begin{enumerate}
    \item Given a trial set of cosmological parameters, we compute the predicted ShapeFit parameters ($\alpha_\textrm{iso}(z_{\rm eff}), \alpha_\textrm{AP}(z_{\rm eff}), f(z_{\rm eff}), \sigma_8(z_{\rm eff}), m+n(z_{\rm eff})$) at each effective redshift, $z_{\rm eff}$.
    \item These predictions are compared to the measured values, and the Gaussian likelihood is computed as 
    \begin{equation}
    \ln \mathcal{L}\propto -\frac{1}{2}(\mathbf{P}_\textrm{meas.}-\mathbf{P}_\textrm{th.})^T\textrm{Cov}^{-1}(\mathbf{P}_\textrm{meas.}-\mathbf{P}_\textrm{th.}),
\end{equation}
    where Cov is the parameter covariance and $\mathbf{P}$ is the vector of parameters, either obtained theoretically or as measured parameters, as in \cite{novellmasot25fullshape}.   
\end{enumerate}
The steps 1 and 2 are iterated within a Markov Chain Monte-Carlo (MCMC) algorithm\footnote{We use the \href{https://emcee.readthedocs.io/en/stable/index.html}{\textsc{emcee}} \cite{Foreman_Mackey_2013} algorithm, in four independent runs, and assess convergence with the rank-normalized $\hat{R}$ statistic \cite{arviz}, computed using the \href{https://python.arviz.org/}{\textsc{ArviZ}} package across the four runs. We require $\hat{R}<1.01$ for all sampled parameters.}, effectively sampling the posterior distribution of cosmological parameters. 

In the baseline $\Lambda$CDM case, our model parameters are the present matter and baryon physical densities, respectively $\omega_m\equiv \Omega_m h^2$ and $\omega_b\equiv \Omega_b h^2$, the Hubble parameter $h\equiv H_0\times 10^{-2} [({\rm km/s Mpc}^{-1})^{-1}]$, the primordial amplitude of fluctuations, expressed as $A_s$, and the spectral index $n_s$. In reporting results we display the matter densities, the Hubble parameter and the amplitude of fluctuations as $\{\Omega_m,\Omega_b,\,H_0,\,\ln(A_s\times10^{10})\}$.

In all cases, the parameters $\omega_m,\,h,\,\ln(A_s\times10^{10})$  are sampled with a uniform improper prior.
We impose a Gaussian prior of $\omega_b=0.02218\pm 0.00055$ following the big bang nucleosynthesis constraints \cite{schoneberg_bbn}. 
LSS data is not informative on baryon density and without a $\omega_b$ prior either from cosmic microwave background (CMB) or BBN the sound horizon scale remains largely unconstrained.

Additionally, we use a broad Gaussian prior for $n_s$, equivalent to ten times the Planck 2018 constraints: $n_{s10}=0.9649\pm 0.042$. The sum of neutrino masses is fixed to $\sum m_\nu = 0.06\, {\rm eV}$ in all cases except in the case where  we explore the constraints on the sum of neutrino masses (see Section \ref{sec:mnu}). For extended models, we vary the additional relevant parameters: e.g., for the $(w_0, w_a)$ model we include those in the parameter vector (with uniform priors); and for free $\sum m_\nu$, we allow it to vary (with a flat prior $\ge 0$).%
All choices of priors are also stated in Table~\ref{tab:priors}.

When combining the DESI clustering measurements with CMB information, we adopt a compressed CMB likelihood defined on the parameter set $\mu_{\rm CMB}\equiv\{\theta_*,\omega_b,\omega_c\}$, where $\theta_*$ is the angular size of the sound horizon at photon decoupling. This likelihood, also used in \cite{desiDR2} to extended cosmological scenarios such as $w_0w_a$CDM, is modelled as a multivariate Gaussian with mean vector $\mu_\textrm{CMB}$ and covariance $\Sigma_\textrm{CMB}$, calibrated from CMB anisotropy measurements following \cite{Lemos23CMB}. The mean and covariance take the following values:

\begin{equation}
\label{eq:cmb}
    \mu_\textrm{CMB} \equiv \begin{pmatrix} \theta_\star \\ \omega_b \\ \omega_c \end{pmatrix} = \begin{pmatrix} 0.01041 \\ 0.02223 \\ 0.14208 \end{pmatrix};\quad \Sigma_\textrm{CMB}=10^{-9}\times
\begin{pmatrix}
0.00662   & 0.12440   & -1.19287 \\
0.12440  & 21.34416  & -94.00083 \\
-1.19287 & -94.00083 & 1488.41710
\end{pmatrix}.
\end{equation}

\begin{table}[ht]
\centering
\begin{tabular}{|llcl|}
\hline
\textbf{Model} & \textbf{Parameter} & \textbf{Default} & \textbf{Prior} \\
\hline
\hline
\multirow{5}{*}{DESI ($\Lambda$CDM)} 
  & $\omega_\textrm{cdm}$ & — & $\mathcal{U}[0.05,\ 0.2]$  \\
  & $\omega_b$ & — & $\mathcal{N}[0.02218,\ 0.00055^2]$  \\
  & $H_0$ [$\mathrm{km\,s^{-1}\,Mpc^{-1}}$] & — & $\mathcal{U}[50,\ 90]$  \\
  & $\ln(10^{10}A_s)$ & — & $\mathcal{U}[2,\ 4]$  \\
  & $n_s$ & — & $\mathcal{N}(0.9649,\ 0.042^2)$  \\
\hline
\multicolumn{4}{|l|}{Beyond $\Lambda$CDM} \\
(sound-horizon marginalization) & $\beta$ & $1$ & $\mathcal{U}[0.5,\ 2]$ \\
(dynamical DE) & $w_0$ & $-1$ & $\mathcal{U}[-3,\ 1]$ \\
 & $w_a$ & $0$ & $\mathcal{U}[-3,\ 2]$ \\
(massive $\nu$) & $\sum m_\nu$ [eV] & $0.06$ & $\mathcal{U}[0,\ 5]$ \\
\hline
\end{tabular}
\caption{Summary of cosmological parameters considered in this work with their prior ranges. For the scenarios beyond $\Lambda$CDM, we also display their default value, which is fixed in the cases where that parameter is not being constrained. For the evolving dark energy equation of state parameters $w_0,w_a$, the condition $w_0+w_a<0$ is imposed, to ensure a period of high-redshift matter domination.}
\label{tab:priors}
\end{table}

\section{Results}\label{sec:results}

\subsection{Baseline \texorpdfstring{$\Lambda$}{L}CDM results}\label{sec:baseline}
\begin{figure}[tbp]
\centering
\includegraphics[width=0.85\textwidth]{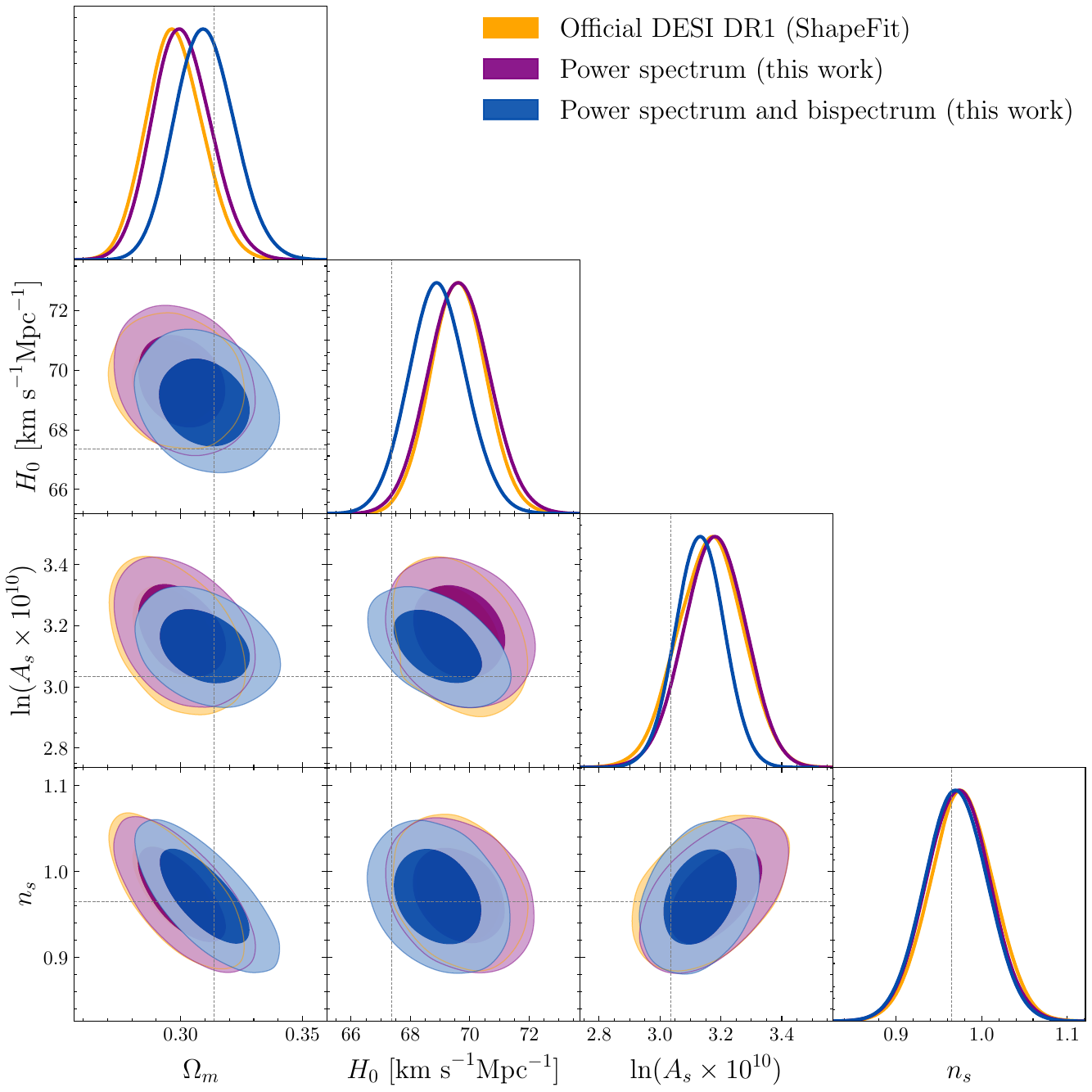}
\caption{\label{fig:PvsB}Baseline $\Lambda$CDM (68 and 95\% confidence levels) joint constraints on the LRG+QSO samples from the DESI DR1 obtained with the ShapeFit compression method, complemented by the BAO postreconstruction data for the BGS, ELGs and Lyman-$\alpha$ tracers. In orange contours, the DESI DR1 official ShapeFit results~\cite{KP5Desi}, in purple the results obtained using the methodology of this work (see Section~\ref{sec:methodology}) when only using the power spectrum, and in blue when the bispectrum is also added. The horizontal and vertical black dashed lines display the Planck 2018 $\Lambda$CDM  central values. We note that including the bispectrum leads to constraints on $\ln(A_s\times10^{10})$ reduced by 20\% with respect to the power spectrum-only approach, and that the power spectrum results from this analysis are completely consistent with the ones from the official DESI DR1 analysis. }%
\end{figure}
Under the baseline $\Lambda$CDM model and for the LRG and QSO samples, the ShapeFit inference from DESI DR1 official measurements \cite{KP5Desi} yields the orange constraints shown in Figure \ref{fig:PvsB}. In that Figure, we also display our results from both the power spectrum-only and the joint power spectrum-bispectrum analysis (respectively in purple and blue) for the same samples, obtained following the methodology described in Section~\ref{sec:methodology}.
The Planck 2018 $\Lambda$CDM  central values are reported as the dotted lines, which are within the 1$\sigma$ region of all posteriors except for $H_0$, which is at $\sim2\sigma$. Interestingly, the addition of the bispectrum shifts the posteriors slightly towards the Planck $\Lambda$CDM results (which was also found in an independent analysis of DESI DR1 that also involved the bispectrum, \cite{philcox25desi}).

The excellent agreement between our power spectrum-only results and the analogous case from the official DESI DR1 results when employing the ShapeFit compression -- considering the changes in modelling and pipeline choices--  highlights the robustness of these results. 

Adding the bispectrum tightens the constraints for the amplitude of the perturbations parameter $\ln(A_s\times 10^{10}$) by $\sim 20\%$ compared to using the power spectrum alone. In particular, 

\begin{align}
    \ln(A_s\times 10^{10}) = \begin{cases}
            3.18\pm 0.10, \text{ for P alone}\\
            3.125\pm 0.079, \text{ for P+B}.
           \end{cases}
\end{align}

The significant reduction in the uncertainty of $\ln(A_s \times 10^{10})$ is achieved through the bispectrum’s role in breaking the degeneracy between $f$ and $\sigma_\textrm{s8}$.
Similar results were also found by Ref~\cite{philcox25desi} in an independent analysis of DESI DR1 data when combining the power spectrum bispectrum signals.

The rest of the central values and 68\% confident levels are summarized in Table \ref{tab:constraints}.

\subsection{Sound-horizon marginalization}\label{sec:shfree}
A key assumption in standard BAO analyses is that the sound horizon at the drag epoch, $r_d$, is known from pre-recombination physics (and usually calibrated through $\omega_b$ informed by Planck or BBN), and is a derived parameter within the $\Lambda$CDM model. Here, we consider the scenario where we drop this assumption, allowing an effective $r_d$ to be a free extra parameter in the fits. Operationally, we follow Ref~\cite{brieden2023tale}:
in each step of the likelihood exploration we multiply the $\Lambda$CDM-inferred $r_d$ parameter by an extra nuisance parameter $\beta$. The ``new'' effective sound horizon scale, $r_d^{\rm eff}=\beta\times r_d^{\Lambda {\rm CDM}}$, is then employed when interpreting the ShapeFit variables such as $\alpha_{\rm iso}$ or $\sigma_{\rm s8}$, where we consider a uniform wide prior for $\beta$ of $\mathcal{U}(0.5,2)$.

This procedure allows a practical marginalisation over $r_d$ within the ShapeFit formalism, which removes the absolute calibration on the BAO physical scale. 
Note, however, that the marginalised BAO scale is still exploited in the parameter inference. We still assume that the same effective $r_d^{\rm eff}$ is shared across redshift bins ($r_d$ is $z$-independent), and that it is an isotropic quantity. This allows us to still employ the BAO scale (with an unknown size) to constrain dimensionless late-time expansion history quantities, such as $\Omega_m$. 

In this setup, the cosmological parameters that rely on absolute distance calibrations, such as $H_0$, do not depend on the BAO scale anchor, and are ``sound-horizon-free''. For the case of a BAO-only analysis, $H_0$ would remain fully unconstrained. However, there is an additional shape information in the no-wiggle power spectrum, through the transfer-function slope, that provides sensitivity to the equality scale within the ShapeFit framework, the matter-radiation equality scale. This scale arises as a characteristic shape of the matter transfer function in the transition between radiation- and matter-dominated epochs. 

Early in the universe’s history, radiation dominance suppressed the growth of matter fluctuations for modes that entered the horizon. Once the universe became matter-dominated, these sub-horizon modes could grow linearly with the scale factor. Because smaller-scale modes entered the horizon earlier (during the radiation era), their growth was delayed relative to larger-scale modes. This results in a transfer function that is nearly flat for large scales ($k<k_{\rm eq}$) and suppressed for small scales ($k>k_{\rm eq}$). This transition defines a characteristic scale --the horizon size at matter-radiation equality-- which appears as the peak of the linear matter power spectrum at $k_{\rm eq}\simeq 0.02\,h{\rm Mpc}^{-1}$.

The exact value of $k_{\rm eq}$ depends on $\Omega_mh^2$ within a late-time $\Lambda$CDM model, and it is related to the shape parameter $m+n$, after a prior on $n_s$ and $\Omega_b h^2$ are given. In this case, we keep the same priors as in the standard $\Lambda$CDM case displayed in Table~\ref{tab:priors}. Thus, $m+n$ constrains $\Omega_mh^2$ calibrated at the equality scale, and the BAO uncalibrated analysis returns $\Omega_m$, which in combination results in a ``sound-horizon-free'' determination of $H_0$ from DESI clustering information alone.

\begin{figure}[htbp]
\centering
\includegraphics[width=0.75\textwidth]{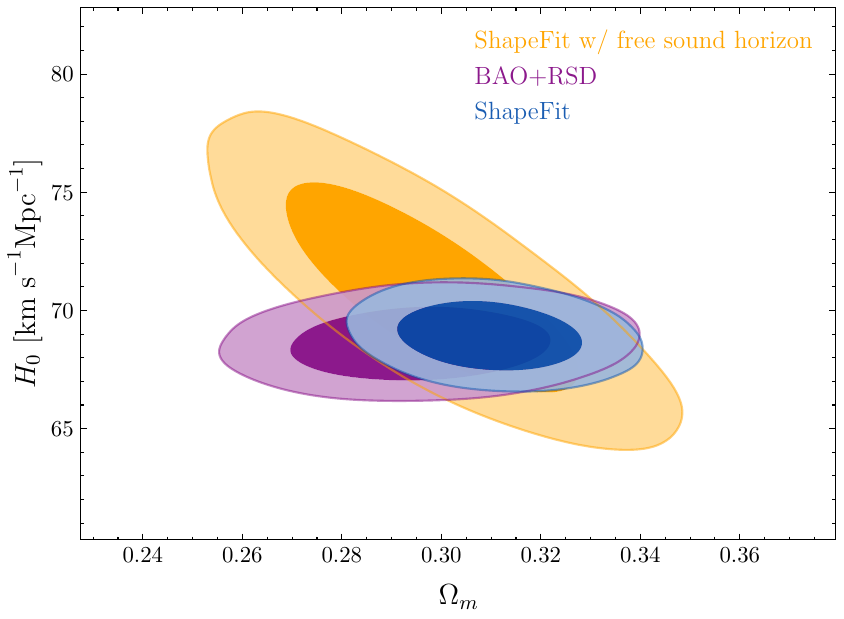}
\caption{\label{fig:shfree}Constraints on $\Omega_m$ and $H_0$ with and without assuming an effective free sound horizon scale parameter $r_d^{\rm eff}$ obtained from the power spectrum and bispectrum ShapeFit results of DESI DR1. The blue contours correspond to the baseline ShapeFit $\Lambda$CDM results employing the standard pre-recombination physics $r_d(w_b,w_c,h,\ldots)$, which is a derived quantity from the model's cosmological parameters. The orange contours feature the same analysis settings, except that the sound horizon scale is effectively treated as a free independent parameter and marginalised over, $r_d^{\rm eff}$. Allowing the horizon scale to freely vary introduces a strong degeneracy between $H_0$ and $\Omega_m$, which follows $\Omega_m h^2$ degeneracy, and that significantly broadens the $H_0$ constraints. For completeness, we display in purple contours a similar analysis to the blue contours, but without using the shape parameters $m+n$ (i.e. the so-called standard compression or BAO+RSD).}
\end{figure}

Figure \ref{fig:shfree} displays the impact of marginalising over the sound horizon scale on the $\Omega_m$–$H_0$ 2D posterior. The blue contours (baseline in this work) assume the standard pre-recombination physics in the sound horizon derivation, $r_d(w_b,w_c,h,\ldots)$, while the orange contours (sound-horizon-free) treat the sound horizon scale as a free effective parameter, $r_d^{\rm eff}$, and marginalise over its value.
In the $r_d$-free case, the contours elongate along the degeneracy $\Omega_m h^2={\rm constant}$, dominated by the ShapeFit $m+n$ parameter. This degeneracy is only broken when the sound horizon anchor is also used, as in the blue contours. Thus, by removing the sound horizon scale anchor, we degrade the precision on the 1-D marginalised $H_0$ parameter, from $\simeq 1.3\%$ up to $\simeq 5\%$,
\begin{align}
    H_0\, [{\rm km\, s^{-1}\,{Mpc}^{-1}}] = \begin{cases}
            70.9\pm 2.9, \text{ sound horizon-free}\\
            68.92\pm 0.97, \text{ sound horizon-anchored}
           \end{cases}
\end{align}
from DESI DR1 clustering alone. The $H_0$ sound horizon-free result is consistent with both the Planck $\Lambda$CDM value and local distance ladder measurements \cite{riess2022comprehensive,casertano25}. %
The sound-horizon free result is also consistent with previous literature on applying this very same methodology to BOSS+eBOSS data \cite{brieden2023tale}, as well as with slightly different approaches for sound-horizon-free determinations of $H_0$ using DESI DR1 data \cite{zaborowski2025sound, zaborowski2025B}. %

\subsection{Evolving dark energy (\texorpdfstring{$w_0w_a$}{wowa})}\label{sec:w0wa}
In this section, we investigate a dynamical dark energy model using the  CPL parametrization $w(a) = w_0 + w_a(1-a)$ \cite{chevallier2001accelerating,lindercpl}, with $w_0$ being the present-day equation-of-state parameter and $w_a$ the parameter describing its evolution with scale factor. This CPL form is a common extension to test whether the data prefer an equation of state different from the cosmological constant value ($w=-1$). Even though it is a phenomenological model, many alternative dark energy scenarios can be expressed in terms of a pair $(w_0,w_a)$. 

\begin{figure}[htbp]
\centering
\includegraphics[width=0.75\textwidth]{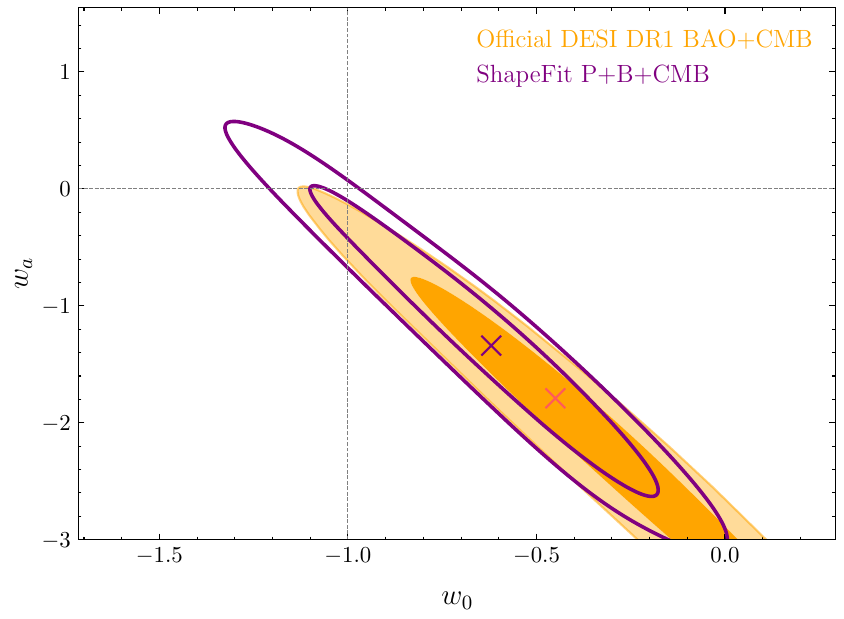}
\caption{Joint 68 and 95\% confidence levels on the dark energy equation-of-state parameters $w_0$ and $w_a$ from the DESI DR1 power spectrum and bispectrum ShapeFit analysis. The crosses indicate the obtained  maximum a posteriori values from the two analyses, and dotted lines indicate the cosmological constant values ($w_0 = -1$, $w_a = 0$). The results are consistent with $\Lambda$CDM at the $2\sigma$ level, with no significant deviation in either parameter. The slight elongation of the baseline contour reflects the fact that most of the constraining power on the dark energy equation of state comes from the expansion history, which is more tightly constrained by post-reconstruction BAO analyses \cite{kp6bao}. The constraints obtained with our methodology (in purple), which use post-reconstruction BAO data for the ELG, BGS and Lyman-$\alpha$ tracers and full-shape information for the LRG and QSO samples, are compared with the official DESI DR1 full post-reconstructed BAO+CMB results from \cite{kp6bao}.}
\label{fig:w0wa}
\end{figure}

In this case, we include in our likelihood an additional term relative to the $\Lambda$CDM analysis, together with an early-universe prior derived from the Planck CMB likelihood, as described by Eq.~\ref{eq:cmb} in Section \ref{sec:methodology}. We allow the parameters $w_0$ and $w_a$ to vary according to their uniform priors, as shown in Table \ref{tab:priors}, and obtain the constraints that are shown in Figure \ref{fig:w0wa}. We display the 2D posteriors for our full-shape DESI DR1 for $w_0w_a$ (in purple), compared with the results obtained from the post-reconstruction DESI DR1 BAO analysis \cite{kp6bao} (in orange), in both cases for the LRG and QSO bins and with the minimal CMB prior, together with the post-reconstructed BAO data from BGS, ELG and Lyman-$\alpha$.

We recover the expected degeneracy direction in the $w_0-w_a$ plane, which arises because the data primarily constrain the effective value of $w(z)$ at the redshift range probed by the observations. As a result, different combinations of $w_0$ and $w_a$ that yield a similar value of $w(z)$ at that redshift produce nearly identical expansion histories and are therefore degenerate.

The DESI BAO measurements themselves are fully consistent with a cosmological constant $\Lambda$ ($w_0=-1,w_a=0$). However, when combined with CMB data they yield a $\sim3\sigma$ preference away from $\Lambda$, a tension that increases if uncalibrated supernovae samples are also added in the data-vector \cite{kp6bao}. The Full Shape (ShapeFit) analysis with the inclusion of the bispectrum presented here, constrains the dark energy equation of state to be compatible at $2\sigma$ with a cosmological constant $\Lambda$: $w_0=-0.62\pm0.29,w_a=-1.34\pm0.82$, finding no significant evidence of dynamical dark energy.

When comparing the recovered errorbars for $w_0$ and $w_a$ reported in this work with the post-reconstructed BAO+CMB constraints found in \cite{kp6bao} (the comparison displayed in Figure \ref{fig:w0wa}), we observe both an increase of the errorbars of $\sim 10\%$ with respect to \cite{kp6bao}, and a shift of $\sim0.6\sigma$ towards $\Lambda$CDM. Both the shift and the elongation of the posteriors of the ShapeFit P+B analysis compared to the BAO ones are in part driven by the effect of reconstruction. The BAO analysis uses reconstruction (which helps tighten the constraints on the expansion history) while the P+B analysis of this work does not use the BAO reconstructed signal for the LRG and QSO samples, as it would require a complete covariance between the ShapeFit and the BAO parameters (see Ref~\cite{forero-sanchez26} as an example on how to account for such full-covariance).

\subsection{Neutrino mass constraints}\label{sec:mnu}
Massive neutrinos leave an imprint on large-scale structure by suppressing clustering on small scales (which can only be measured in full-shape analyses) and altering the expansion rate (which alters the cosmological distances and therefore has an effect on the BAO). In the baseline analysis, we held the sum of neutrino masses at the value of $\sum m_\nu = 0.06$ eV, and in this Section we allow $\sum m_\nu\,[{\rm eV}]$ to vary as a free parameter across the uniform prior $\mathcal{U}[0,5]$. 

Because neutrino mass reduces the amplitude of the power spectrum at small scales (i.e. high $k$), we expect ShapeFit to be sensitive to it primarily through the shape parameter $m+n$. Still, since the neutrino density was more relevant at early epochs of the expansion history, most of the information on $\sum m_\nu$ is present in pre-recombination probes such as the CMB \cite{kp6bao}. We therefore compare the results from DESI DR1 alone with the combination of DESI with the compressed Planck likelihood presented in Section~\ref{sec:methodology}.

\begin{figure}[tbp]
\centering
\includegraphics[width=0.47\textwidth]{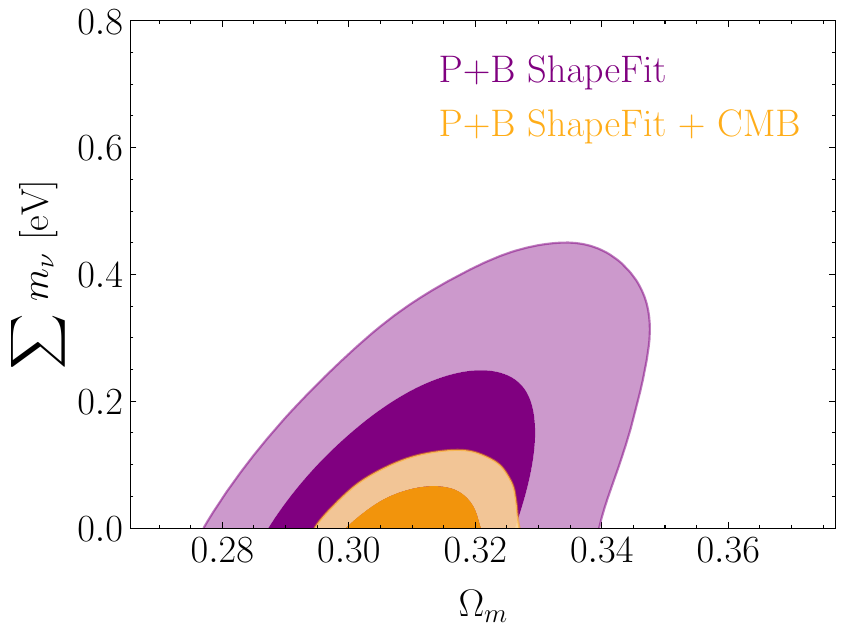}
\includegraphics[width=0.43\textwidth]{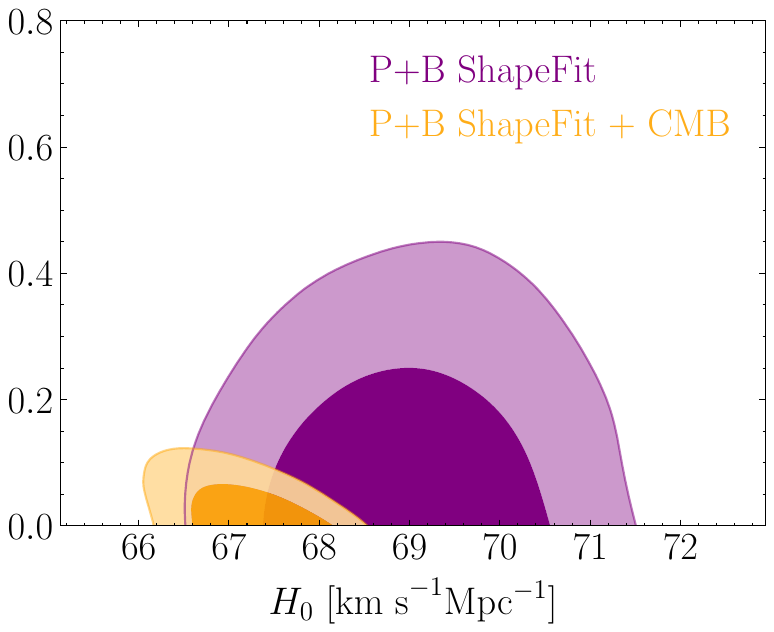}
\caption{\label{fig:mnu}Marginalized posterior for the sum of neutrino masses $\sum m_\nu$ from the DESI DR1 ShapeFit analysis (in purple) together with the constraints of DESI DR1 combined with the compressed Planck likelihood (in orange). This is consistent with previous results, both from the CMB \cite{collaboration2018planck} and the DESI collaboration \cite{kp6bao,KP7}, which favour a sum below 100 meV.}%
\end{figure}

Figure \ref{fig:mnu} shows the 2D joint posteriors for $\sum m_\nu$  with the parameters $\Omega_m$ and $H_0$, respectively. As is common in LSS results, the constraints for $\sum m_\nu$ hit the physical $\sum m_\nu=0$ prior. We obtain upper limits of respectively,
\begin{itemize}
    \item $\sum m_\nu < 0.36$ eV (95\% DESI DR1),
    \item $\sum m_\nu < 0.10$ eV (95\% DESI DR1+CMB),
\end{itemize}
 which can also be found in Table \ref{tab:constraints}. As expected, DESI DR1 data alone are not sufficiently sensitive to discriminate between neutrino mass hierarchies\footnote{We find that when we perform the same analysis without the addition of the bispectrum, $\sum m_\nu<0.39$ eV (95\%), so the inclusion of the bispectrum yields a slight improvement of $\sim8\%$, when not combining with CMB data.  }. However, when combined with the compressed Planck likelihood, the inverted hierarchy (which requires $\sum m_\nu \gtrsim 0.1$ eV) is disfavoured at approximately the 95\% confidence level, whereas the normal hierarchy (with $\sum m_\nu \sim 0.06$ eV) remains consistent with the best-fit results. Our upper bound is comparable to those obtained in previous Planck+LSS analyses, which report limits of order $\sim 0.07$ eV \cite{kp6bao,KP7}.

\section{Discussion}\label{sec:discussion}

\begin{table}[htbp]
\centering
        \small
    \resizebox{\columnwidth}{!}{
\begin{tabular}{|l|c|c|c|c|}
\hline 
Parameter & $\Lambda$CDM & Free $r_d$ & $w_0$-$w_a$ (+CMB) & $\sum m_\nu$ (+CMB) \\
\hline 
\hline
$\Omega_m$ & $0.310\pm0.012$ & $0.298\pm 0.020$ & $0.334\pm0.027$ & $0.3119\pm0.0064$ \\
$H_0$ [km s$^{-1}$ Mpc$^{-1}$] & $68.92\pm0.97$ & $70.9\pm 2.9$ & $65.3\pm2.7$ & $67.13\pm0.50$ \\
$\ln(A_s\times10^{10})$ & $3.125\pm0.079$ & unconstrained & $3.185\pm0.094$ & $3.207\pm0.069$ \\
$w_0$ & $-1^\dag$ & $-1^\dag$ & $-0.62\pm0.29$ & $-1^\dag$ \\
$w_a$ & $0^\dag$ & $0^\dag$ & $-1.34\pm0.82$ & $0^\dag$  \\
$\sum m_\nu$ [eV] & $0.06^\dag$ & $0.06^\dag$ & $0.06^\dag$ & $<0.10$ (95\% CL) \\
\hline
\end{tabular}}
\caption{\label{tab:constraints}Summary of cosmological parameter constraints (maximum a posteriori values and $1\sigma$ uncertainties) from DESI DR1 ShapeFit analyses for the joint power spectrum multipoles and bispectrum monopole data-vector, under various scenarios. The columns correspond to: {\it i)} the baseline $\Lambda$CDM model; {\it ii)} a $\Lambda$CDM-like scenario with free sound horizon, $r_d$; {\it iii)} a dynamical dark energy model with free $w_0$ and $w_a$; and {\it iv)} a model with the total neutrino mass parameter, $\sum m_\nu$, free. In these last two cases, we also add CMB data in the form of the $\mu_{\rm CMB}$ data-vector of Equation~\ref{eq:cmb}, noted as +CMB. The $\Lambda$CDM and Free $r_\textrm{d}$ cases use Gaussian priors on $\omega_\textrm{b}$ from BBN \cite{schoneberg_bbn}, and in all cases we consider priors for $n_\textrm{s}$ corresponding to 10 times the Planck 2018 constraints \cite{collaboration2018planck}, as explained in Section \ref{sec:methodology}.  For parameters that are not varied in a given scenario, we indicate the fixed value without errorbars with a $^\dag$ symbol (e.g., $w_0=-1^\dag$ for baseline $\Lambda$CDM). CL stands for confidence levels. }
\end{table}

We have presented the cosmological constraints from the DESI DR1  joint power spectrum and bispectrum galaxy clustering data, for both standard $\Lambda$CDM and extensions varying the dark energy equation of state and the neutrino mass. We have used the ShapeFit results obtained in \cite{novellmasot25fullshape} from the joint power spectrum and bispectrum data-vector for LRG and QSO tracers, combining them with the reconstructed BAO constraints for BGS and ELG \cite{kp3bao}, as well as the BAO for the Lyman-$\alpha$ forest from \cite{kp4lya}. A summary of results can be found in Table \ref{tab:constraints}, and we highlight the main points in what follows.

For the baseline $\Lambda$CDM, we find best-fit parameter values broadly in agreement with both Planck CMB results \cite{collaboration2018planck} and the DESI DR1 Full-Shape analysis of the same data \cite{KP5Desi,KP7}: both the matter density $\Omega_m$ and amplitude of perturbations $\ln(A_s\times10^{10})$ are compatible at $1\sigma$ with both Planck and DESI DR1 full-shape results. Interestingly, the addition of the bispectrum in the data vector causes the recovered constraints on $\Omega_m$ and $H_0$ to lie closer to Planck than when using only the power spectrum.
    
The inclusion of the bispectrum in our joint analysis significantly improves the sensitivity to the growth of structure, with errorbars on $\ln(A_s\times10^{10})$ being reduced by roughly 20\% when the joint data-vector is used. On this parameter alone, adding the bispectrum in the analysis yields equivalent precision improvement as the statistical power gained by using $\sim1.8$ times more data.

To further assess the robustness of our $H_0$ constraints, we also performed an analysis where the sound horizon at the drag epoch, $r_d$, is treated as a free parameter, following the procedure of \cite{brieden2023tale}. This effectively decouples the low-redshift clustering information from assumptions about early-Universe physics, so marginalizing over $r_d$ allows a more model-independent inference of the Hubble constant. As expected, this leads to a broader posterior, with $H_0 = 71.68 \pm 3.44$ km $\textrm{s}^{-1}\textrm{Mpc}^{-1}$, compared to $H_0 = 68.92 \pm 0.98$ km$\textrm{s}^{-1}\textrm{Mpc}^{-1}$ in the baseline $\Lambda$CDM case with fixed $r_d$. The central value shifts upward, while increasing errorbars with respect to \cite{brieden2023tale,zaborowski2025sound}.

Beyond the baseline model, we explored several extensions (combining with a compressed CMB likelihood), in all cases using the tracer and data-vector choices stated in Section \ref{sec:methodology}, as used also in the $\Lambda$CDM case\footnote{This yields broader constraints than other results that combine both full-shape and post-reconstruction BAO data for all tracers. In this work, we opt for showing the results without any combination of analyses on any given tracer, and without adding Supernovae data.}. Allowing the dark energy equation of state parametrized by $w_0$–$w_a$, the DESI ShapeFit constraints remain consistent with a cosmological constant. In this parametrization, combining our clustering results with CMB data yields $w_0 \approx -0.6\pm0.3$, $w_a \approx -1.3\pm0.8$, which is within ~$2\sigma$ of  $\Lambda\equiv \{w_0=-1,,w_a=0\}$. The observed trend --the preference for $w_0 > -1$ and $w_a < 0$, primarily driven by the observed distance measurements in the LRG bins-- is similar to that reported in the official DESI BAO and full-shape analyses \cite{kp6bao,KP7}, but is not yet statistically significant in our case. 

When allowing the sum of neutrino masses to vary, our ShapeFit analysis (with a compressed \textit{Planck} prior) finds no evidence for a non-zero mass: the posterior is consistent with the minimal mass scenario within normal hierarchy ($\sum m_\nu \approx 0.06$ eV). The inverted hierarchy (which requires $\sum m_\nu \gtrsim 0.1$ eV) is disfavoured at $\sim95\%$ confidence by our analysis of DESI DR1 data combined with CMB information. We obtain an upper limit of $\sum m_\nu < 0.1$ eV (95\% confidence limits) from ShapeFit, which is slightly weaker than the limit from the DESI full-shape analysis, $\sum m_\nu < 0.071$ eV at 95\% confidence limits \cite{KP7}. The difference can be explained as follows. Firstly, we do not combine full-shape and BAO data in each tracer as we do not dispose of their cross-covariance terms. Neutrino mass weakly affects the expansion history (parametrized by the BAO); additionally, some small-scale clustering information (important for neutrino mass sensitivity) may not be captured by the ShapeFit compression as fully as with the Full Modelling approach.

Overall, across all considered extensions (including time-varying dark energy and massive neutrinos), our analysis shows no strong evidence of new physics beyond $\Lambda$CDM, with some results currently being in 2$\sigma$ tension with Planck.

It is instructive to compare our results with those from the more detailed Full Modelling approaches, specifically with the state-of-the-art DESI Collaboration’s Full Modelling analysis of the same DR1 data \cite{KP7}. We find agreement between the $\Lambda$CDM constraints presented here and the ones presented in \cite{KP5Desi}. For example, the Full Modelling analysis finds %
$\{\Omega_m,\,H_0 [{\rm km} {\rm s}^{-1} {\rm Mpc}^{-1}],\,\ln(A_s\times10^{10})\}=\{0.284\pm0.010,\, 70.0\pm1.0,\,3.10\pm0.10\}$, while we find $\{0.310\pm0.012,\, 68.92\pm0.98,\,3.125\pm0.079\}$. 

Our ShapeFit uncertainties (excluding the growth of structure parameter, which is better constrained due to the addition of the bispectrum) are  $\sim 10-30\%$ larger for parameters such as $H_0$, the $w_0w_a$ parametrization or $\sum m_\nu$. These differences in error bars are to be expected: in the work presented in this paper, we do not combine the full-shape and the reconstructed BAO information, as done, e.g. in \cite{KP5Desi}. Such a combination would yield the expansion history and growth of structure to be maximally constrained, but this goes beyond the scope of this paper. Additionally, by construction, ShapeFit compresses the broadband power spectrum information into a few parameters, with a slight loss of information compared to fitting the full shape. This loss is insignificant in the case of $\Lambda$CDM, as shown in \cite{KP5Desi}, but can become relevant for extensions of $\Lambda$CDM such as the one presented in Section \ref{sec:w0wa}. 

Most importantly, we do not observe any significant bias between Full Modelling analysis on the power spectrum-only data-vector and the Shapefit analysis using both power spectrum and bispectrum: all constraints obtained from  ShapeFit are statistically consistent with those from the Full Modelling pipeline, not only for the base $\Lambda$CDM parameters but also for the extensions. Both analyses, for instance, report the same qualitative preference for $w_0>-1$, $w_a<0$ when DESI is combined with CMB data (Figure \ref{fig:w0wa}), and both find neutrino masses to be consistent with the normal hierarchy and constrained to the sub-0.1 eV level.

The ShapeFit methodology has proven to be a powerful yet simple tool for performing cosmological inference in a model-independent way. Its compressed parameter set captures the main cosmological information via the BAO parameters $\alpha_\textrm{iso},\alpha_\textrm{AP}$, the growth of structure parameters $f,\sigma_\textrm{s8}$, and the shape parameter $m+n$. This comes at the cost of marginally weaker constraints in some beyond-$\Lambda$CDM scenarios (due to not directly fitting the full power spectrum), but the simplicity and modularity of ShapeFit have other advantages: one can easily introduce alternate parameters or perform consistency tests (such as freeing the sound-horizon scale $r_d$ to test late-time vs. early-time physics) without re-running a complete model for the power spectra. This is especially useful given that analysing the power spectra or bispectra directly is an increasingly complex task, needing to account for many sources of systematics (see e.g. \cite{KP5Desi,novellmasot25fullshape,gsponer2025fiducial,findlay2024exploring,pinon2025mitigation}). Using the ShapeFit compression in a similar way as presented in this paper allows cosmologists, both inside and outside the DESI collaboration, to obtain robust cosmological constraints in a straight-forward way. Moreover, the availability of the intermediate ShapeFit parameters provides a summary of the clustering information that is complementary to Full Modelling approaches, and can help diagnose the origin of possible differences between datasets or between low- and high-redshift probes, particularly in the context of current cosmological tensions (e.g. \cite{verde2019tensions}).

Looking ahead, the DESI survey is expected to increase the survey volume by almost an order of magnitude %
\cite{besuner2025spectroscopic}.
The resulting reduction in statistical errors will require refining the methodology to ensure theoretical systematics remain subdominant.
Necessary future improvements include modelling the window function for the bispectrum, implementing the $\theta$-cut method for mitigating fibre assignment systematics \cite{pinon2025mitigation} and updating the strategy to extract the shape parameter $m$ \cite{ghaemi2025smooth}.

In conclusion, in this work, we have presented the first joint power spectrum and bispectrum DESI DR1 cosmology results using ShapeFit. While these results do not suggest growing tensions with respect to previous works \cite{KP7}, %
the constraints are expected to improve significantly as more data are collected. The tools and results shown here can serve as a foundation for those future analyses.

\section*{Data Availability}
All data from the tables and figures are publicly available in machine-readable format at \href{https://doi.org/10.5281/zenodo.19099550}{10.5281/zenodo.19099550}, in compliance with the DESI
data management plan.
\acknowledgments
SNM, HGM and LV thank Hernán Noriega, Nicola Deiosso, Uendert Andradea, Benjamin Weaver and Alex Krolewski for the helpful discussion and suggestions.
SNM acknowledges funding from the official doctoral programme of the University of Barcelona for the development of a research project under the PREDOCS-UB grant.
HGM acknowledges support through the Leonardo programme (LEO23-1-897) of the BBVA foundation and through the programmes Ram\'on y Cajal (RYC-2021-034104) and Consolidación Investigadora (CNS2023-144605) of the Spanish Ministry of Science and Innovation. 

Funding for this work was partially provided by the Spanish MINECO under project  PID2022-141125NB-I00 MCIN/AEI, and the `Center of Excellence Maria de Maeztu 2020-2023' award to the ICCUB (CEX2019-000918-M \sloppy funded by MCIN/AEI/10.13039/501100011033).

This material is based upon work supported by the U.S. Department of Energy (DOE), Office of Science, Office of High-Energy Physics, under Contract No. DE–AC02–05CH11231, and by the National Energy Research Scientific Computing Center, a DOE Office of Science User Facility under the same contract. Additional support for DESI was provided by the U.S. National Science Foundation (NSF), Division of Astronomical Sciences under Contract No. AST-0950945 to the NSF’s National Optical-Infrared Astronomy Research Laboratory; the Science and Technology Facilities Council of the United Kingdom; the Gordon and Betty Moore Foundation; the Heising-Simons Foundation; the French Alternative Energies and Atomic Energy Commission (CEA); the National Council of Humanities, Science and Technology of Mexico (CONAHCYT); the Ministry of Science, Innovation and Universities of Spain (MICIU/AEI/10.13039/501100011033), and by the DESI Member Institutions: \url{https://www.desi.lbl.gov/collaborating-institutions}. Any opinions, findings, and conclusions or recommendations expressed in this material are those of the author(s) and do not necessarily reflect the views of the U. S. National Science Foundation, the U. S. Department of Energy, or any of the listed funding agencies.

The authors are honored to be permitted to conduct scientific research on I'oligam Du'ag (Kitt Peak), a mountain with particular significance to the Tohono O’odham Nation.

This work has made use of the following publicly available codes:  \href{https://emcee.readthedocs.io/en/stable/index.html}{\textsc{Emcee}} \cite{Foreman_Mackey_2013}, \href{https://numpy.org/}{\textsc{NumPy}} \cite{numpy}, \href{https://getdist.readthedocs.io/en/latest/}{\textsc{GetDist}} \cite{getdist}, \href{https://matplotlib.org}{\textsc{Matplotlib}} \cite{matplotlib}, \href{astropy.org}{\textsc{Astropy}} \cite{astropy}, \href{http://class-code.net/}{\textsc{Class} \cite{class}}. We are grateful to the developers who made these codes public.

\appendix

\section{The ShapeFit Interpretation Vademecum}\label{sec:appB}
In this section, we include a short description on how to interpret the ShapeFit compressed variables in terms of cosmological variables. All  this information can be found in existing literature \cite{brieden2021shapefit,brieden2022model,brieden2023tale}, and we refer the reader to those references  for further details. 

The dilation parameters, $\alpha_\perp(z)$ and $\alpha_\parallel(z)$ are related to the angular diameter distance, $D_M(z)$, the Hubble distance, $D_H(z)$, respectively, and the sound horizon scale at drag epoch, $r_d$ such that,
\begin{eqnarray}
 \label{eq:alpha1}   \alpha_\parallel(z)&\equiv&\frac{D_H(z)/r_d}{[D_H(z)/r_d]^{\rm ref}},\\
\label{eq:alpha2}    \alpha_\perp(z)&\equiv&\frac{D_M(z)/r_d}{[D_M(z)/r_d]^{\rm ref}},
\end{eqnarray}
where ref stands for the reference (or fiducial) cosmology at which the redshift-to-distance conversion is performed, and the reference cosmology of the linear power spectrum template. These distances are defined from the Hubble expansion history, $H(z)$, such that,
\begin{eqnarray}
 D_H(z)&\equiv &\frac{c}{H(z)},\\
    D_M(z)&\equiv& \frac{c}{H_0\sqrt{\Omega_k}}\sinh{\left[\sqrt{\Omega_k}\int_0^z\frac{dz'}{H(z')/H_0}\right]},
\end{eqnarray}
where $H_0$ is the expansion rate today, $H_0\equiv H(z=0)$, and $\Omega_k$ is the `curvature density' parameter.
The sound horizon scale at the drag epoch, $r_d$ depends on  pre-recombination parameters, such as the expansion history before recombination, the sound horizon speed in the baryon-photon fluid, and on the redshift of decoupling (or more precisely, the drag epoch, when the baryons stop feeling the presence of the photons). Note how by determining the best-fitting $\alpha_\parallel$ and $\alpha_\perp$ we can only measure distances in terms of a reference scale: the sound horizon $r_d$. This is a key point when defining the quantity $\sigma_{\rm s8}$ below.

The shape  parameters $m$ and $n$ parametrize variations of the linear matter template reference such that,
\begin{equation}
P_{\rm lin}(k)=P_{\rm lin}^{\rm ref}(k) \left\{ \frac{m}{a_{\rm SF}}\tanh \left[ a_{\rm SF}\ln\left(\frac{k}{k_p}\right) \right] + n \ln\left(\frac{k}{k_p}\right)  \right\},
\end{equation}
where the values for $a_{\rm SF}$ and $k_p$ are chosen to be, $a_{\rm SF}=0.6$ and $k_p=\pi/r_d^{\rm ref}$ (although other values could also be chosen).
On one hand, the $m$ parameter is the slope at the pivot scale $k_p$. Within the $\Lambda$CDM model, it is related to the $\Omega_mh^2$ parameter, containing information on pre-recombination physics. On the other hand, the $n$ parameter controls the power-law index of the power spectrum at large $k$ values and is related to any effective primordial spectral index, such as the parameter $n_s$ in $\Lambda$CDM.
In terms of connecting these parameters to the cosmological parameters, we can write the ShapeFit parameter $m$ as \cite{brieden2021shapefit,noriega2024comparing},
\begin{equation}
    m=\frac{d}{d\ln k}\left\{ \ln\left[\frac{P_{\rm lin,\,nw}(k/s)/P_\mathcal{R}(k)
}{P^{\rm ref}_{\rm lin,\,nw}(k/s)/P^{\rm ref}_\mathcal{R}(k)
}\right]\right\}_{k=k_p},
\end{equation}
where the subscript `nw' stands for non-wiggle, $P_{\rm lin,\, nw}$ the linear power spectrum with the BAO wiggles filtered out\footnote{See Appendix B of \cite{brieden2022model} as well as \cite{ghaemi2025smooth} for a detailed comparison of different smoothing prescriptions.}, $s$ is the sound horizon scale in units of its reference value, $s\equiv r_d/r_d^{\rm ref}$, and $P_\mathcal{R}(k)$ is the (power law) primordial power spectrum, $P_\mathcal{R}(k)=A_s(k/k_p)^{n_s-1}$, with $A_s$ the primordial amplitude of perturbations and $n_s$ the primordial spectral index. The connection to the second shape parameter, $n$ in this case, is simply,
\begin{equation}
    n=n_s-n_s^{\rm ref}.
\end{equation}
Note that the relation above may differ for cosmologies with non-zero running of the spectral index.

The (logarithmic) growth factor $f$ stands for the logarithmic rate of change of the linear growth parameter $D(a)$ with respect to the logarithm of the scale factor, $a$,
\begin{equation}
    f(z)\equiv\left.\frac{d\ln{D(z')}}{d\ln{a(z')}}\right|_{z'=z}.
\end{equation}

The amplitude of dark matter fluctuations filtered by an $8 \,{\rm Mpc}h^{-1}$ scale is usually defined as, 
\begin{equation}
\sigma_8^2(z)\equiv\int_0^\infty d^3k P_{\rm lin}(k,z)W_{\rm TH}(k\cdot 8{\rm Mpc}h^{-1}),
\end{equation}
where $P_{\rm lin}$ the linear matter power spectrum, and $W_{\rm TH}(k\cdot R)$ a top-hat filter with a scale of $R=8\,{\rm Mpc}h^{-1}$ in this case. However, this quantity is not what it is measured by any fixed-template measurement, such as ShapeFit or the traditional compressed method. The traditional compression method (this is the ShapeFit compression where $m$ and $n$ are kept fixed to 0) implicitly assumes an “early-time rescaling”, which induces a change in the interpretation of $\sigma_8$ via $\alpha_\parallel$ and $\alpha_\perp$. Therefore, the actual smoothing scale used to define $\sigma_8$ in the equation above is not kept fixed while exploring parameter space during fixed-template fits.

Instead, ShapeFit measures the power spectrum amplitude at a reference scale in units of the sound horizon $r_d$ (recall how the distances $D_M$ and $D_H$ are measured {\it in terms of} $r_d$ in Eqs~\ref{eq:alpha1}-\ref{eq:alpha2}).  Additionally, upon changing the shape parameters $m$ and $n$,  the power spectrum amplitude remains fixed only at the pivot scale $k_p = \pi/r_d^{\rm ref}$; the shape of the power spectrum does change, which in turns affects the $\sigma_8$ value. Therefore, in reality, ShapeFit does not constrain $\sigma_8$ but effectively constrains the variable $A_{sp}^{1/2}$, where $A_{\rm sp}$ is the power spectrum amplitude evaluated at $k_p/s$, which is the quantity that is kept fixed during the parameter exploration,
\begin{equation}
    A_{\rm sp}=\frac{1}{s^3}P_{\rm lin,\,nw}(k_p/s).
\end{equation}
Note the overall scaling with $1/s^3$, which appears to describe the volume change when all scales are rescaled by a factor of $s$ (see section 3.1 of \cite{brieden2021shapefit} for a detailed explanation).

The issue with the definition of $\sigma_8$ (i.e., the smoothing scale varying along with other parameters) can easily be accounted for by defining the fluctuation amplitude in such a way that it does not change during the fitting process, i.e., such that it is independent of changes in $s$, $8\, {\rm Mpc}h^{-1}\rightarrow s\cdot 8\,{\rm Mpc}h^{-1}$
\begin{equation}
    \sigma_{s8}^2\equiv \int_0^\infty d^3k P_{\rm lin}(k,z)W_{\rm TH}(k\cdot s\cdot 8{\rm Mpc}h^{-1}).
\end{equation}
Relating $A_{\rm sp}$ to $\sigma_{\rm s8}$ via $\left(\sigma_{\rm s8}/\sigma_{\rm s8}^{\rm ref}\right)^2=A_{\rm sp}/A_{\rm sp}^{\rm ref}$  ignores the changes in the shape of $P_{\rm lin}$ (beyond its overall amplitude) related to the $m$ and $n$ parameters.
The exact value of $\sigma_{s8}$ requires integration over $P_{\rm lin}(k|m,n)$ at each step of the MCMC sampling of the posterior. %
A good approximation is given in 
\cite{brieden2023tale},
\begin{equation}
    \left(\frac{\sigma_{\rm s8}}{\sigma_{\rm s8^{\rm ref}}}\right)^2\simeq\frac{A_{\rm sp}}{A_{sp}^{\rm ref}}\exp\left\{ \frac{m+n}{a_{\rm SF}}\tanh \left[a_{\rm SF}\ln\left(\frac{r_d^{\rm ref}[{\rm Mpc}h^{-1}]}{8\, {\rm Mpc} h^{-1}}\right) \right]\right\}.
\end{equation}
Note again that ShapeFit can internally work with the parameter $A_{\rm sp}$ (or $f A_{\rm sp}^{1/2}$) and perform the cosmological interpretation without having to derive $\sigma_{\rm s8}$. The only reason to derive $\sigma_{\rm s8}$ is historical, as to provide a quantity which is similar to the traditional $\sigma_8$ parameter, but with a 
definition that remains consistent and transparent during parameter exploration.

\section{Sequential ShapeFit compression}\label{sec:compression}

\begin{figure}[htbp]
\centering
\includegraphics[width=0.85\textwidth]{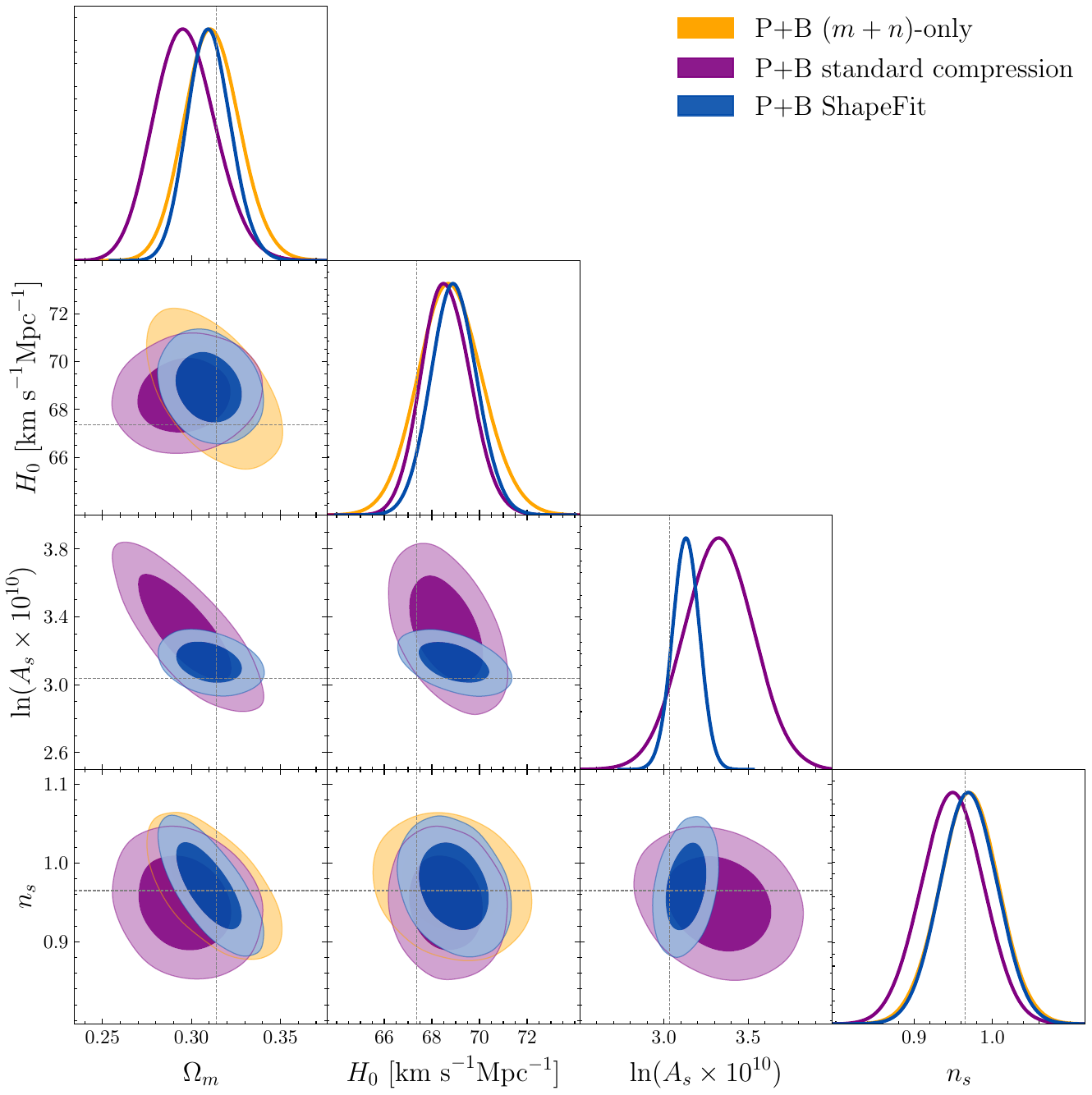}
\caption{\label{fig:compression}Results for different subsets of the ShapeFit compression. We show marginalized posterior distributions for the $\Lambda$CDM parameters $\Omega_m$, $H_0$, $\ln(A_s\times10^{10})$ inferred from the DESI DR1 data using the joint power spectrum-bispectrum analysis. The blue contours correspond to the full ShapeFit analysis (which are therefore identical to the blue constraints in Figure \ref{fig:PvsB}). The orange contours use only information from the combined shape parameter $m+n$, while the purple constraints are obtained from the standard compression approach (using the parameters $\alpha_\parallel,\alpha_\bot,f,\sigma_\textrm{s8}$).}
\end{figure}

 We perform the cosmological inference following Section \ref{sec:methodology} and assuming a $\Lambda$CDM model with priors defined in Table \ref{tab:priors},  using different subsets of the ShapeFit parameters. First, we analyze the case where only the combined shape parameter $m+n$ is used for the cosmological parameter estimation (shown in yellow in Figure~\ref{fig:compression}). We then derive  (Figure~\ref{fig:compression} in purple) the constraints for its complementary, standard compression, subset consisting of $\alpha_\textrm{iso}, \alpha_\textrm{AP},f,\sigma_\textrm{s8}$. These results are compared with those obtained using the full ShapeFit framework, which includes both the standard compression parameters and $m+n$ (blue in Figure~\ref{fig:compression}). In all cases, we combine the ShapeFit constraints on the LRG and QSO tracers with the post-reconstructed BAO from BGS, ELG and Lyman-$\alpha$, as done in all results in the present paper.

As expected, the full ShapeFit constraints lie approximately at the intersection of the other two cases. 
Although the shape parameter $m+n$ contains no direct information on the growth of structure, its inclusion significantly reduces the uncertainties in $\ln(A_s\times10^{10})$ as it breaks the strong degeneracy between $\Omega_m$ and $\ln(A_s\times10^{10})$ present in the standard compression.

\input{authorlist.affiliations}
\bibliographystyle{JHEP}
\bibliography{main}

\end{document}

%% file: authorlist.tex
\affiliation{Affiliations are in Appendix \ref{sec:affiliations}}

\author[1,2]{{S.~Novell-Masot},}
\author[3,4,1]{{H.~Gil-Mar\'in}\orcidlink{0000-0003-0265-6217},}
\author[5,1]{{L.~Verde}\orcidlink{0000-0003-2601-8770},}
\author[6]{{J.~Aguilar},}
\author[7]{{S.~Ahlen}\orcidlink{0000-0001-6098-7247},}
\author[8,9]{{D.~Bianchi}\orcidlink{0000-0001-9712-0006},}
\author[10]{{D.~Brooks},}
\author[4,11]{{F.~J.~Castander}\orcidlink{0000-0001-7316-4573},}
\author[6]{{T.~Claybaugh},}
\author[12]{{S.~Cole}\orcidlink{0000-0002-5954-7903},}
\author[13]{{A.~de la Macorra}\orcidlink{0000-0002-1769-1640},}
\author[14]{{J.~Della~Costa}\orcidlink{0000-0003-0928-2000},}
\author[6,15]{{S.~Ferraro}\orcidlink{0000-0003-4992-7854},}
\author[5,16]{{A.~Font-Ribera}\orcidlink{0000-0002-3033-7312},}
\author[17,18]{{J.~E.~Forero-Romero}\orcidlink{0000-0002-2890-3725},}
\author[4,19,11]{{E.~Gaztañaga}\orcidlink{0000-0001-9632-0815},}
\author[20]{{S.~Gontcho A Gontcho}\orcidlink{0000-0003-3142-233X},}
\author[21]{{A.~X.~Gonzalez-Morales}\orcidlink{0000-0003-4089-6924},}
\author[22]{{G.~Gutierrez},}
\author[6]{{J.~Guy}\orcidlink{0000-0001-9822-6793},}
\author[23]{{C.~Hahn}\orcidlink{0000-0003-1197-0902},}
\author[24,25]{{H.~K.~Herrera-Alcantar}\orcidlink{0000-0002-9136-9609},}
\author[26,27,28]{{K.~Honscheid}\orcidlink{0000-0002-6550-2023},}
\author[29]{{C.~Howlett}\orcidlink{0000-0002-1081-9410},}
\author[30]{{M.~Ishak}\orcidlink{0000-0002-6024-466X},}
\author[16]{{J.~Jimenez}\orcidlink{0000-0001-8528-3473},}
\author[14]{{R.~Joyce}\orcidlink{0000-0003-0201-5241},}
\author[31]{{R.~Kehoe},}
\author[32]{{D.~Kirkby}\orcidlink{0000-0002-8828-5463},}
\author[6]{{A.~Kremin}\orcidlink{0000-0001-6356-7424},}
\author[28]{{C.~Lamman}\orcidlink{0000-0002-6731-9329},}
\author[33]{{L.~Le~Guillou}\orcidlink{0000-0001-7178-8868},}
\author[34,16]{{M.~Manera}\orcidlink{0000-0003-4962-8934},}
\author[14]{{A.~Meisner}\orcidlink{0000-0002-1125-7384},}
\author[5,16]{{R.~Miquel},}
\author[19]{{S.~Nadathur}\orcidlink{0000-0001-9070-3102},}
\author[21,35]{{G.~Niz}\orcidlink{0000-0002-1544-8946},}
\author[36,37,38]{{W.~J.~Percival}\orcidlink{0000-0002-0644-5727},}
\author[39]{{I.~P\'erez-R\`afols}\orcidlink{0000-0001-6979-0125},}
\author[40]{{G.~Rossi},}
\author[41,42,43]{{L.~Samushia}\orcidlink{0000-0002-1609-5687},}
\author[44]{{E.~Sanchez}\orcidlink{0000-0002-9646-8198},}
\author[45]{{E.~F.~Schlafly}\orcidlink{0000-0002-3569-7421},}
\author[6]{{D.~Schlegel},}
\author[46,47]{{M.~Schubnell},}
\author[6]{{J.~Silber}\orcidlink{0000-0002-3461-0320},}
\author[14]{{D.~Sprayberry},}
\author[47]{{G.~Tarl\'{e}}\orcidlink{0000-0003-1704-0781},}
\author[14]{{B.~A.~Weaver},}
\author[48]{{C.~Zhao}\orcidlink{0000-0002-1991-7295},}
\author[6]{{R.~Zhou}\orcidlink{0000-0001-5381-4372},}

%% file: authorlist.affiliations.tex
\section{Author Affiliations}
\label{sec:affiliations}

\noindent \hangindent=.5cm $^{1}${Institut de Ci\`encies del Cosmos (ICCUB), Universitat de Barcelona (UB), c. Mart\'i i Franqu\`es, 1, 08028 Barcelona, Spain.}

\noindent \hangindent=.5cm $^{2}${Institut Qu\'imic de Sarri\`a-Universitat Ramon Llull (IQS-URL), Via Augusta 390, 08017, Barcelona, Spain}

\noindent \hangindent=.5cm $^{3}${Departament de F\'{\i}sica Qu\`{a}ntica i Astrof\'{\i}sica, Universitat de Barcelona, Mart\'{\i} i Franqu\`{e}s 1, E08028 Barcelona, Spain}

\noindent \hangindent=.5cm $^{4}${Institut d'Estudis Espacials de Catalunya (IEEC), c/ Esteve Terradas 1, Edifici RDIT, Campus PMT-UPC, 08860 Castelldefels, Spain}

\noindent \hangindent=.5cm $^{5}${Instituci\'{o} Catalana de Recerca i Estudis Avan\c{c}ats, Passeig de Llu\'{\i}s Companys, 23, 08010 Barcelona, Spain}

\noindent \hangindent=.5cm $^{6}${Lawrence Berkeley National Laboratory, 1 Cyclotron Road, Berkeley, CA 94720, USA}

\noindent \hangindent=.5cm $^{7}${Department of Physics, Boston University, 590 Commonwealth Avenue, Boston, MA 02215 USA}

\noindent \hangindent=.5cm $^{8}${Dipartimento di Fisica ``Aldo Pontremoli'', Universit\`a degli Studi di Milano, Via Celoria 16, I-20133 Milano, Italy}

\noindent \hangindent=.5cm $^{9}${INAF-Osservatorio Astronomico di Brera, Via Brera 28, 20122 Milano, Italy}

\noindent \hangindent=.5cm $^{10}${Department of Physics \& Astronomy, University College London, Gower Street, London, WC1E 6BT, UK}

\noindent \hangindent=.5cm $^{11}${Institute of Space Sciences, ICE-CSIC, Campus UAB, Carrer de Can Magrans s/n, 08913 Bellaterra, Barcelona, Spain}

\noindent \hangindent=.5cm $^{12}${Institute for Computational Cosmology, Department of Physics, Durham University, South Road, Durham DH1 3LE, UK}

\noindent \hangindent=.5cm $^{13}${Instituto de F\'{\i}sica, Universidad Nacional Aut\'{o}noma de M\'{e}xico,  Circuito de la Investigaci\'{o}n Cient\'{\i}fica, Ciudad Universitaria, Cd. de M\'{e}xico  C.~P.~04510,  M\'{e}xico}

\noindent \hangindent=.5cm $^{14}${NSF NOIRLab, 950 N. Cherry Ave., Tucson, AZ 85719, USA}

\noindent \hangindent=.5cm $^{15}${University of California, Berkeley, 110 Sproul Hall \#5800 Berkeley, CA 94720, USA}

\noindent \hangindent=.5cm $^{16}${Institut de F\'{i}sica d’Altes Energies (IFAE), The Barcelona Institute of Science and Technology, Edifici Cn, Campus UAB, 08193, Bellaterra (Barcelona), Spain}

\noindent \hangindent=.5cm $^{17}${Departamento de F\'isica, Universidad de los Andes, Cra. 1 No. 18A-10, Edificio Ip, CP 111711, Bogot\'a, Colombia}

\noindent \hangindent=.5cm $^{18}${Observatorio Astron\'omico, Universidad de los Andes, Cra. 1 No. 18A-10, Edificio H, CP 111711 Bogot\'a, Colombia}

\noindent \hangindent=.5cm $^{19}${Institute of Cosmology and Gravitation, University of Portsmouth, Dennis Sciama Building, Portsmouth, PO1 3FX, UK}

\noindent \hangindent=.5cm $^{20}${University of Virginia, Department of Astronomy, Charlottesville, VA 22904, USA}

\noindent \hangindent=.5cm $^{21}${Departamento de F\'{\i}sica, DCI-Campus Le\'{o}n, Universidad de Guanajuato, Loma del Bosque 103, Le\'{o}n, Guanajuato C.~P.~37150, M\'{e}xico}

\noindent \hangindent=.5cm $^{22}${Fermi National Accelerator Laboratory, PO Box 500, Batavia, IL 60510, USA}

\noindent \hangindent=.5cm $^{23}${Department of Astronomy, The University of Texas at Austin, 2515 Speedway Boulevard, Austin, TX 78712, USA}

\noindent \hangindent=.5cm $^{24}${Institut d'Astrophysique de Paris. 98 bis boulevard Arago. 75014 Paris, France}

\noindent \hangindent=.5cm $^{25}${IRFU, CEA, Universit\'{e} Paris-Saclay, F-91191 Gif-sur-Yvette, France}

\noindent \hangindent=.5cm $^{26}${Center for Cosmology and AstroParticle Physics, The Ohio State University, 191 West Woodruff Avenue, Columbus, OH 43210, USA}

\noindent \hangindent=.5cm $^{27}${Department of Physics, The Ohio State University, 191 West Woodruff Avenue, Columbus, OH 43210, USA}

\noindent \hangindent=.5cm $^{28}${The Ohio State University, Columbus, 43210 OH, USA}

\noindent \hangindent=.5cm $^{29}${School of Mathematics and Physics, University of Queensland, Brisbane, QLD 4072, Australia}

\noindent \hangindent=.5cm $^{30}${Department of Physics, The University of Texas at Dallas, 800 W. Campbell Rd., Richardson, TX 75080, USA}

\noindent \hangindent=.5cm $^{31}${Department of Physics, Southern Methodist University, 3215 Daniel Avenue, Dallas, TX 75275, USA}

\noindent \hangindent=.5cm $^{32}${Department of Physics and Astronomy, University of California, Irvine, 92697, USA}

\noindent \hangindent=.5cm $^{33}${Sorbonne Universit\'{e}, CNRS/IN2P3, Laboratoire de Physique Nucl\'{e}aire et de Hautes Energies (LPNHE), FR-75005 Paris, France}

\noindent \hangindent=.5cm $^{34}${Departament de F\'{i}sica, Serra H\'{u}nter, Universitat Aut\`{o}noma de Barcelona, 08193 Bellaterra (Barcelona), Spain}

\noindent \hangindent=.5cm $^{35}${Instituto Avanzado de Cosmolog\'{\i}a A.~C., San Marcos 11 - Atenas 202. Magdalena Contreras. Ciudad de M\'{e}xico C.~P.~10720, M\'{e}xico}

\noindent \hangindent=.5cm $^{36}${Department of Physics and Astronomy, University of Waterloo, 200 University Ave W, Waterloo, ON N2L 3G1, Canada}

\noindent \hangindent=.5cm $^{37}${Perimeter Institute for Theoretical Physics, 31 Caroline St. North, Waterloo, ON N2L 2Y5, Canada}

\noindent \hangindent=.5cm $^{38}${Waterloo Centre for Astrophysics, University of Waterloo, 200 University Ave W, Waterloo, ON N2L 3G1, Canada}

\noindent \hangindent=.5cm $^{39}${Departament de F\'isica, EEBE, Universitat Polit\`ecnica de Catalunya, c/Eduard Maristany 10, 08930 Barcelona, Spain}

\noindent \hangindent=.5cm $^{40}${Department of Physics and Astronomy, Sejong University, 209 Neungdong-ro, Gwangjin-gu, Seoul 05006, Republic of Korea}

\noindent \hangindent=.5cm $^{41}${Abastumani Astrophysical Observatory, Tbilisi, GE-0179, Georgia}

\noindent \hangindent=.5cm $^{42}${Department of Physics, Kansas State University, 116 Cardwell Hall, Manhattan, KS 66506, USA}

\noindent \hangindent=.5cm $^{43}${Faculty of Natural Sciences and Medicine, Ilia State University, 0194 Tbilisi, Georgia}

\noindent \hangindent=.5cm $^{44}${CIEMAT, Avenida Complutense 40, E-28040 Madrid, Spain}

\noindent \hangindent=.5cm $^{45}${Space Telescope Science Institute, 3700 San Martin Drive, Baltimore, MD 21218, USA}

\noindent \hangindent=.5cm $^{46}${Department of Physics, University of Michigan, 450 Church Street, Ann Arbor, MI 48109, USA}

\noindent \hangindent=.5cm $^{47}${University of Michigan, 500 S. State Street, Ann Arbor, MI 48109, USA}

\noindent \hangindent=.5cm $^{48}${Department of Astronomy, Tsinghua University, 30 Shuangqing Road, Haidian District, Beijing, China, 100190}